# Kagome Flatbands for Coherent Exciton-Polariton Lasing

T. H. Harder,[1,*] O. A. Egorov,[2] C. Krause,[1] J. Beierlein,[1] P. Gagel,[1] M. Emmerling,[1] C. Schneider, [1,3] U. Peschel,[2] S. Höfling[1,4], and S. Klembt[1,#]

[1]Technische Physik, Wilhelm-Conrad-Röntgen-Research Center for Complex Material Systems, and Würzburg-Dresden Cluster of Excellence ct.qmat, Universität Würzburg, Am Hubland, D-97074 Würzburg, Germany

[2]Institute of Condensed Matter Theory and Solid State Optics, Abbe Center of Photonics, Friedrich-Schiller-Universität Jena, Max-Wien-Platz 1, D-07743 Jena, Germany

[3]Institute of Physics, University of Oldenburg, D-26129 Oldenburg, Germany

[4]SUPA, School of Physics and Astronomy, University of St. Andrews, KY 16 9SS, United Kingdom

[*]tristan.harder@uni-wuerzburg.de

[#]sebastian.klembt@uni-wuerzburg.de

**Abstract**

Kagome lattices supporting Dirac cone and flatband dispersions are well known as a highly frustrated, two-dimensional lattice system. Particularly the flatbands therein are attracting continuous interest based on their link to topological order, correlations and frustration. In this work, we realize coupled microcavity implementations of Kagome lattices hosting exciton-polariton quantum fluids of light. We demonstrate precise control over the dispersiveness of the flatband as well as selective condensation of exciton-polaritons into the flatband.

Subsequently, we focus on the spatial and temporal coherence properties of the laser-like emission from these polariton condensates that are closely connected to the flatband nature of the system. Notably, we find a drastic increase in coherence time due to the localization of flatband condensates. Our work illustrates the outstanding suitability of the exciton-polariton system for detailed studies of flatband states as a platform for microlaser arrays in compact localized states, including strong interactions, topology and non-linearity.

**Introduction**

The Kagome lattice, or basketweave lattice, that was first studied in the context of magnetism and spin frustration [1] and has led to significant progress in the field of spin liquids [2-7], essentially describes a two-dimensional, pyrochlor-like structure. The band structure of the Kagome lattice features Dirac points, enabling the study of Dirac physics [8-10], as well as a dispersionless band, referred to as a flatband, in which states are localized purely due to the geometry of the lattice. This flatband has raised significant interest due to its non-trivial topological properties [11-18]. Interestingly, signatures of Dirac cones as well as flatbands have very recently been measured in the antiferromagnetic Kagome lattice compound FeSn [19-21]. Following investigations in electronic systems, Kagome lattices have been realized in the context of artificial matter [22] in a range of experimental platforms such as cold atoms [23], optically induced lattices [24], plasmonics [25], and microwave photonic crystals [26]. Additionally, photonic realizations of Kagome flatbands were developed and have spurred research towards flatband lasers [27]. However, so far an experimental implementation of the latter is elusive.

Exciton-polaritons (polaritons) are quasi-particles that arise due to the strong coupling of cavity photons to quantum well excitons [28]. The unique part-light, part-matter composition of these polaritons facilitated the observation of polariton condensates [29] that emit coherent laser light [30]. Using a variety of confinement techniques, lattice potential environments can be designed for exciton-polaritons allowing to study the emergence of polaritonic band structures [31,32] which were subsequently even driven into a topologically non-trivial regime [33,34]. So far, polaritonic flatbands have been realized in Lieb lattices [35,36] where precise control over the flatness of the flatbands has been demonstrated [37], yet no signature of a Kagome flatband has been demonstrated so far [38]. Theoretical proposals, however, have predicted a variety of intriguing effects such as spontaneous symmetry breaking, topologically protected edge states, and edge state solitons due to the interacting, driven-dissipative nature of polaritons [39-41].

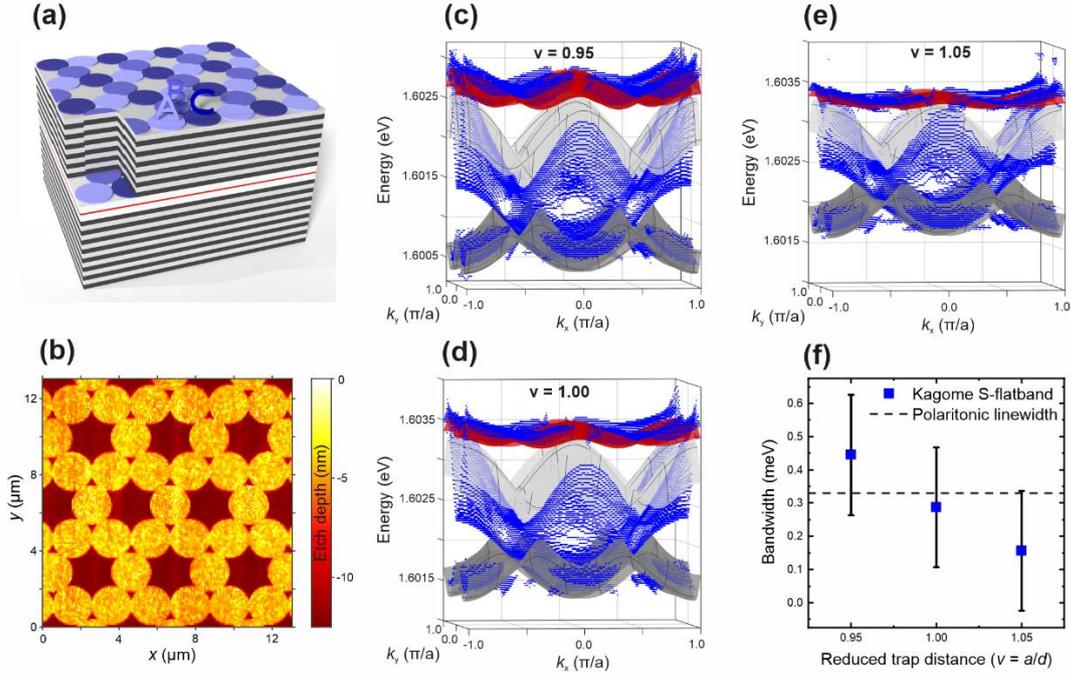

**Figure 1. (a)** Schematic representation of a Kagome lattice implemented in a polaritonic microcavity using the etch-and-overgrowth approach where the confinement is achieved by a local elongation of the cavity layer at the position of the traps. The three sites of the Kagome unit cell are marked by *A*, *B* and *C*. **(b)** Atomic force microscopy image of the structured cavity layer before the top distributed Bragg mirror is grown. **(c)-(e)** Band structure reconstructions from hyperspectral imaging performed on three Kagome lattices with reduced trap distances of v=a/d=0.95, 1.00 and 1.05, respectively, represented by blue data points. The tight-binding model includes nearest and next-nearest neighbor coupling, *t* and *t'* and the flatband is highlighted in red. **(d)** Flatband bandwidth with respect to the reduced trap distance revealing that a bandwidth well below the polariton linewidth was achieved.

**Fabrication of polaritonic Kagome lattices**

The main challenge in realizing a Kagome lattice of coupled semiconductor microresonator pillars, in contrast to for example Lieb or honeycomb lattices, is the dramatic difference in hole sizes that need to be etched in the hexagonal benzene-like structure compared to the much smaller hole in the triangular unit (see Figs. 1 (a), (b)). Due to this requirement, it is exceedingly difficult to reach coupling conditions that are reasonably well described by tight-binding models. In order to circumvent this limitation, we realize a polaritonic Kagome lattice using an etch-and-overgrowth approach. An AlAs microcavity containing two stacks of four 7 nm thick GaAs quantum wells is embedded between two distributed Bragg mirrors (DBR) consisting of 32 (37) $Al_{0.20}Ga_{0.80}As$/AlAs mirror pairs in the top (bottom) mirror. The lattice potential environment is defined using the etch-and-overgrowth approach where the epitaxial growth is interrupted after growing the cavity layer and the traps are defined by etching the surrounding area to a depth of approximately 10 nm. Finally, the top DBR is grown. This local elongation of the cavity at the position of the traps leads to a confinement potential of approximately 6.7 meV. The final cavity is characterized by a vacuum Rabi splitting of 11.4 meV and a quality factor of ~7,400. Due to the high degree of tunability of the confinement potential as well as the coupling strength between the traps, this approach offers striking technological control over band structure properties [37]. A schematic representation of a Kagome lattice in an etch-and-overgrowth microcavity is depicted in Fig. 1(a) and an atomic force microscopy image of the cavity surface acquired before growing the top DBR is presented in Fig. 1(b). Here, it is clearly visible that the small holes in the triangular unit with a diameter of ~500 nm display the same etch depth of ~10 nm as all other etched regions.

**Optical spectroscopy of polaritonic Kagome flatbands**

We start by demonstrating the existence of flatbands in three polaritonic Kagome lattices characterized by trap diameters of $d = 2.0$ μm and reduced trap distances $v = a/d$, with $a$ denoting the center-to-center distance of adjacent traps, of 0.95, 1.00 and 1.05 at a moderate exciton-photon detuning of approximately -5.5 meV. For the optical excitation, a continuous wave laser spot expanded to a diameter of ~15 μm at a power far below condensation threshold was selected (see Supplementary Section for further details). By scanning the Fourier-space image across the entrance slit of the spectrometer and reconstructing the bands from the dataset (c.f. [34,37]), we obtain the full band structure represented by the blue points in Figs. 1(c)-(e). The data are fitted using a tight-binding model that considers nearest and next-nearest neighbor coupling $t$ and $t'$, respectively. The flatband is highlighted in red. As the formation of a flatband relies on the destructive interference on nearest neighbors, deviations from the simplest tight-binding model, such as next-nearest neighbor coupling or mode hybridization, induce a residual dispersiveness of the flatband. The strength of the next-nearest neighbor coupling with respect to the nearest neighbor coupling can be finely tuned by increasing the reduced trap distance $v$, as is depicted in Fig. 1(f), where the bandwidth of the flatband is plotted with respect to the reduced trap distance. At $v = 1.05$, the bandwidth of the flatband is reduced to 150 μeV, which is well below the polariton linewidth of 330 μeV. At the same time, for all reduced trap distances $v$, the well-known dispersion featuring six Dirac-cones is visible below the flatband, indicating relevant inter-site coupling in all cases.

We continue by studying polariton condensation into the *S*-flatband of Kagome lattices. In order to describe the condensation in Kagome lattices theoretically, we use a modified Gross-Pitaevskii (GP) approach [42,43] with realistic sample parameters (see Supporting Information for details). Within this driven-dissipative model, the permanent decay of polaritons from the condensate as well as the reservoir is compensated by an external, off-resonant, time-independent o ptical pump. In order to account for polariton dynamics within the deep potential landscape, we scale the key system parameters, such as stimulated scattering and polariton-polariton interaction, in accordance with the local fraction of the excitonic component by rescaling them with the Hopfield coefficients. In Figs. 2(a) and 2(b), calculated radiation spectra of a Kagome lattice characterized by $v = 1.00$ under incoherent excitation below and above condensation threshold are displayed. The radiation from the S-flatband is clearly visible. Furthermore, at the condensation threshold, the laser emission originates exclusively from Bloch modes of the flatband (Fig. 2(b)). After theoretically predicting selective condensation into the flatband, we experimentally realize a polariton condensate in a lattice with trap diameters of $d = 2.0$ μm and a reduced trap distance of $v = 1.00$. The Fourier space spectra below and above the condensation threshold that are presented in Figs. 2(c) and (d), respectively, unequivocally demonstrate selective lasing emission from the Kagome S-flatband. Furthermore, the input-output characteristic presented in Fig. 2(e) features the distinctive increase in emission intensity of around five orders of magnitude across threshold accompanied by a sudden drop in linewidth below 100 μeV at the threshold power $P_{th}$ that indicates a temporal coherence buildup. Additionally, a blue shift of the emission energy due to increasing polariton-polariton interaction is observed (Fig. 2(f)). The corresponding real-space mode pattern of the polariton Kagome flatband laser is depicted in Fig. 2(g).

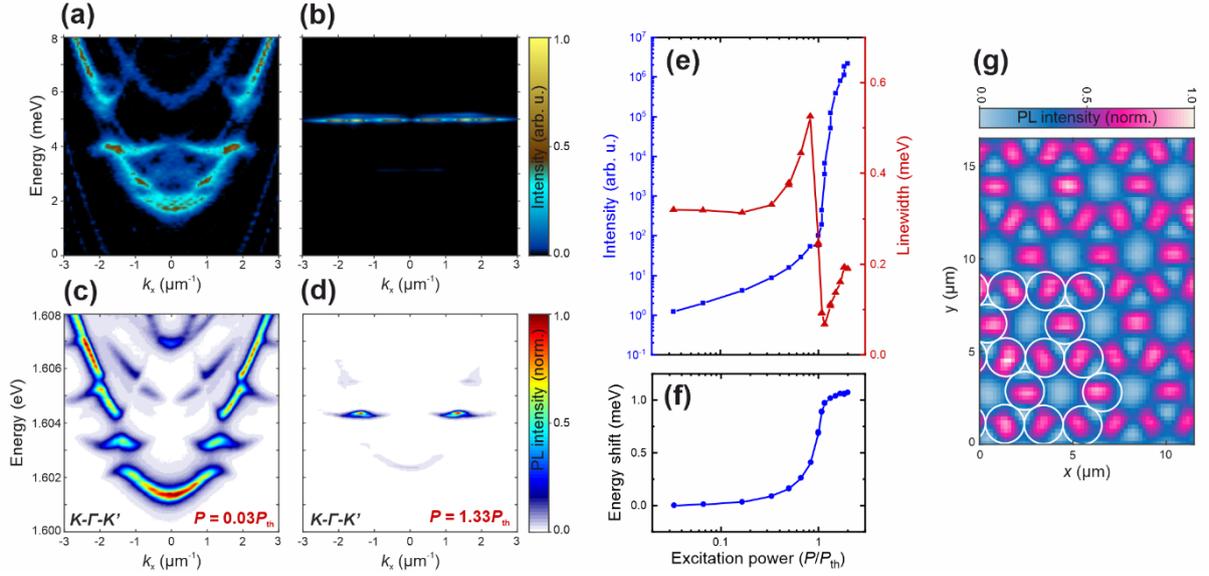

**Figure 2. (a)-(b)** Theoretical population of exciton-polaritons in Fourier space calculated within a Kagome lattice with a reduced trap distance of $v = 1.00$ under an incoherent localized excitation below (a) and above (b) condensation threshold. Condensation clearly occurs in the Kagome flatband. **(c)-(d)** Exemplary experimental Fourier space spectra at excitation powers of $P = 0.03P_{th}$ and $P = 1.33P_{th}$, respectively, revealing the selective excitation of a flatband condensate. **(e)-(f)** Input-output characteristics of a Kagome lattice with trap diameters of $d = 2\mu m$ and a reduced trap distance of $v = 1.00$ featuring the characteristic increase in PL intensity accompanied by a drop in linewidth at the threshold power $P_{th} = 36$ mW of polariton condensation and a continuous blue shift of the emission energy. **(g)** Spectrally integrated real space mode distribution above threshold showing the flatband condensate.

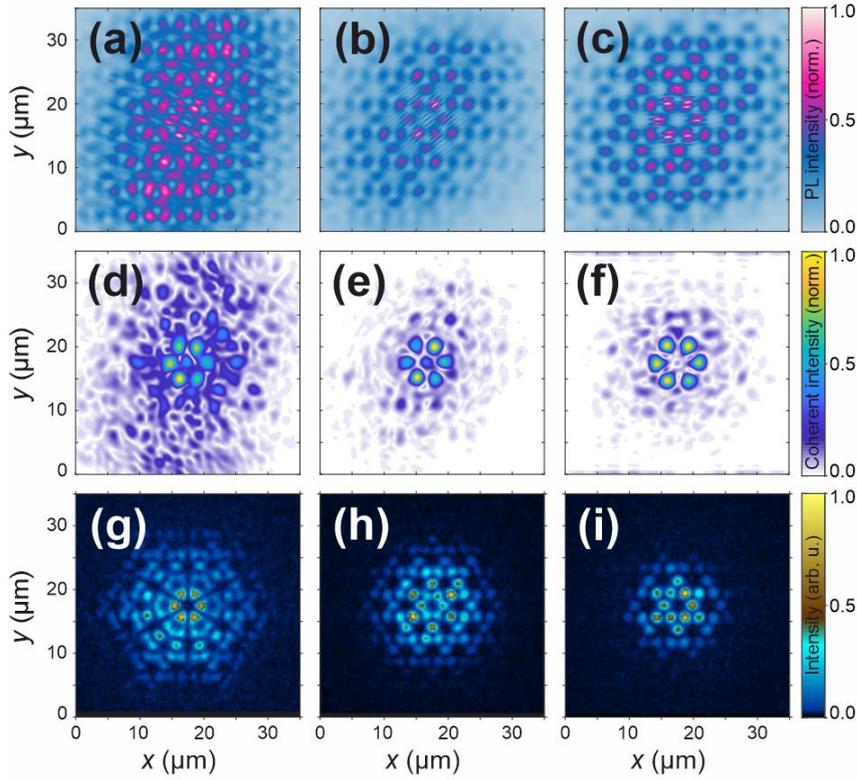

**Figure 3.** The coherence of the flatband condensate of three Kagome lattices with lattice aspect ratios of $v = 0.95$, 1.00 and 1.05 leads to fringe patterns in the Michelson interferograms displayed in **(a)-(c)**, respectively. **(d)-(f)** The experimental coherence length is visualized by plotting only the coherent part of the PL emission. **(g)-(i)** Calculated profiles of flatband condensates under an incoherent excitation for different separations between lattice sites: v=0.91 (g), v=1.00 (h), v=1.11(i).

Having established Kagome flatbands as well as selective polariton condensation into these flatbands, we extensively study the coherence properties of polariton condensates in the three Kagome lattices introduced in Fig. 1 using a range of interferometry techniques. First, we analyze Michelson interferometry measurements at zero delay $g^{(1)}(r, -r, t = 0)$, where one reflector flips the image around the x- and y-axis (c.f. [29]). The extent of the resulting fringe pattern displayed in Figs. 3(a)-(c) is a direct measure of the length scale on which the condensate emits coherently. By Fourier transforming the image, selecting the coherent part and applying an inverse Fourier transformation, this spatial decay of coherence is visualized in Figs. 3(d)-(f). As demonstrated in Fig. 1, the flatband becomes less dispersive and hence closer to a perfectly flat band with increasing reduced trap distance *v*. This decrease in dispersiveness furthermore results in a higher degree of localization of the polaritons which leads to a shorter coherence length. For a lattice with a reduced trap distance of *v* = 1.05, characterized by an almost dispersionless flatband, the coherent emission is localized to a single compact cell of the Kagome lattice, referred to as a compact localized state (CLS).

We furthermore performed numerical simulations of the flatband condensation under incoherent localized pumping for different separations between traps and compared the respective real space profiles (see Figs. 3(g)-3(i)). For the case of a very tight lattice with overlapping sites, the flatband becomes slightly dispersive due to deformation of the traps' eigenmodes, leading to a substantial spatial extension of the condensate in real space presented in Fig. 3(g). By increasing the separation between lattice sites, the localization of the condensate in the flatband is enhanced (see Fig. 3(i)), in full agreement with the experimental findings. Interestingly, while there is flatband lasing from a large area in space, this localization of coherent emission from a CLS, essentially forming a lattice of flatband lasers, is exclusively driven by geometrical frustration.

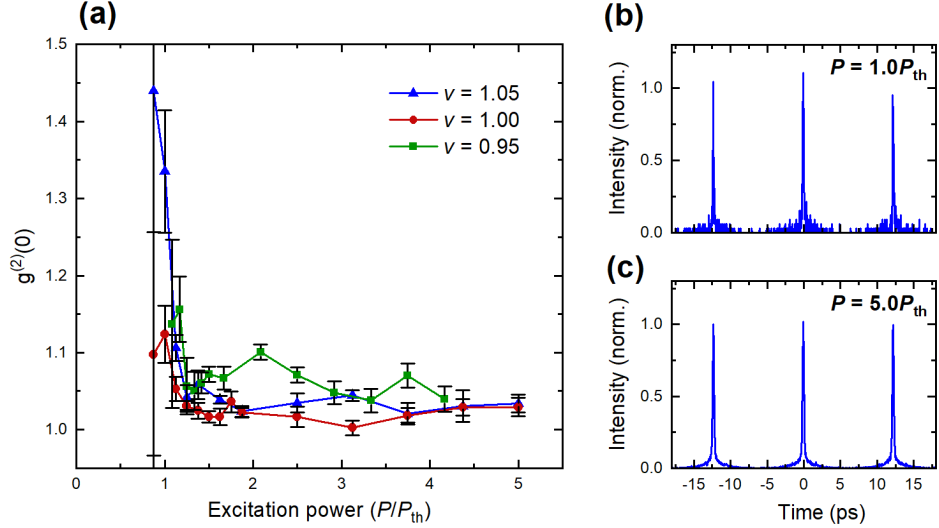

**Figure 4.** (a) Hanbury Brown-Twiss measurement of the second order temporal coherence $g^{(2)}(\tau = 0)$ as a function of the laser excitation power, normalized by the threshold power, for reduced trap distances of v = 0.95, 1.00 and 1.05, respectively. At the threshold ($P/P_{th} = 1$), the second order coherence abruptly drops from a thermal to a coherent lasing state, yielding $g^{(2)}_{v=0.95} \sim 1.06$ and $g^{(2)}_{v=1.00-1.05} \sim 1.03$. In (b) and (c), exemplary second-order coherence properties of the single-mode flatband polariton laser (v=1.05) below and above the threshold for pulsed excitation, respectively, are presented. Here, intensity denotes the coincidence counts from the two avalanche photodiodes.

Next, we study polariton condensates in single CLSs in detail. One of the key properties of a polariton condensate is its coherence that results in lasing emission and can be accessed experimentally by performing Hanbury Brown and Twiss (HBT) interferometry. In Fig. 4(a), an HBT measurement of a CLS polariton condensate which was excited with a small laser spot with a diameter of ~2 µm, focused to the center of the CLS, is presented. Here, a pulsed Ti:sapphire laser tuned to the first Bragg minimum of the stopband and featuring a pulse width of 10 ps that is shorter than the nominal resolution of 40 ps of the single photon detectors was used. Notably, the finite length of the emission pulse acts as an effective time filter, since only photons emitted from the reservoir within the relaxation time are correlated [44]. From these measurements, we are thus able to extract the second order coherence $g^{(2)}(\tau = 0)$ as a function of the pump power and lattice spacing, which is directly related to the dispersiveness of the flatband. An HBT measurement of the second order coherence $g^{(2)}(\tau = 0)$ as a function of the laser excitation power, normalized by the condensation threshold power, for reduced trap distances of $v = 0.95$, 1.00 and 1.05, respectively, is presented in Fig. 4 (a), with respective coincidence data below and above threshold shown in (b) and (c), respectively. At the condensation threshold ($P/P_{th} = 1$), the second order coherence abruptly drops from a thermal to a coherent lasing state, yielding excellent coherence values of $g^{(2)}_{v=0.95} \sim 1.06$ and $g^{(2)}_{v=1.00-1.05} \sim 1.03$ for excitation powers exceeding $P > P_{th}$ and unequivocally demonstrating a highly coherent flatband laser state. The slight decrease of $g^{(2)}$ for larger lattice spacings and thus decreased flatband curvature and increased localization is consistent with previous findings of improved coherence in etched polaritonic micropillars which is attributed to a strongly reduced condensate depletion in confined structures [44]. However, it should be pointed out that in flatbands the localization is governed by geometrical frustration rather than pillar etching, leading to reduced charge noise [45], since considerably less sidewall damage is inflicted.

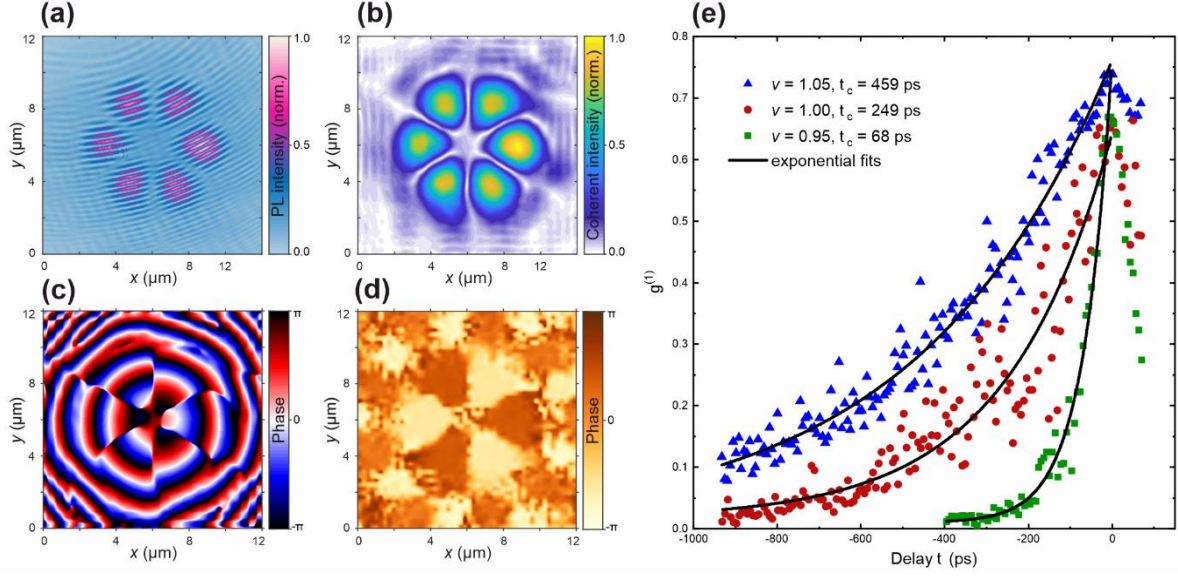

**Figure 5. (a)** Mach-Zehnder interferogram of a CLS condensate in a lattice with a reduced trap distance of $v = 1.05$. **(b)-(c)** Spatial maps of coherence and phase, respectively, extracted from the interferogram. **(d)** The phase profile of the condensate calculated within the GP approach. **(e)** First order coherence function $g^{(1)}$ as a function of delay $\tau$. The exponential fits yield coherence times of $\tau_c = 68$, 249 and 459 ps for lattices with reduced trap distances of $v = 0.95$, 1.00 and 1.05, respectively.

After having established a coherent flatband laser state, we continue by studying the first order spatial as well as temporal coherence properties of the lasing emission. In Fig. 5(a), a Mach-Zehnder interferogram of a CLS polariton condensate which was excited with a small, continuous wave laser spot with a diameter of ~2 µm, focused to the center of the CLS, is presented. One of the six lobes of the CLS condensate was magnified and superimposed with the full CLS condensate emission to serve as a fixed phase reference. The spatial coherence as well as phase maps that are presented in Figs. 5(b)-(c), respectively, were obtained by the Fourier analysis method introduced above and reveal a phase shift of exactly $\pi$ between adjacent lobes of the CLS condensate. Using GP model simulations, we are able to reproduce precisely this phase pattern as shown in Fig. 5(d). This finding is of great importance because this phase shift intuitively visualizes the mechanism that forms a flatband. As polaritons originating from adjacent sites will interfere destructively on their shared nearest neighbor site, propagation is inhibited leading to the formation of the flatbands we observe.

To investigate the temporal behavior of the coherence of such a localized state, we perform Michelson interferometry on a CLS condensate that was excited non-resonantly with a continuous wave (cw) laser. By moving the mirror of one of the arms of the Michelson interferometer, the delay between the images that are superimposed on the camera is tuned. The first order correlation $g^{(1)}$ as a function of the delay $\tau$ is subsequently extracted from the fringe visibility of a line spectrum across one of the lobes, which for the interference of two beam profiles $I_1(x)$ and $I_2(x)$ is given by

$$I(x,\tau) = I_1(x) + I_2(x) + 2|g^{(1)}(\tau)|\sqrt{I_1(x)I_2(x)} \cos\left(\frac{2\pi\theta}{\lambda_0}x + \phi\right)$$

where $\lambda_0$ denotes the wavelength of the condensate and $\theta$ and $\phi$ correspond to the phase difference and angle between the beam profiles [30]. The time-dependent first order correlation $g^{(1)}(\tau)$ including exponential fits of the coherence decay for the three lattices introduced above is plotted in Fig. 5(e). As expected, with decreasing dispersiveness of the flatband and consequently increasing localization of the CLS, the coherence time drastically increases from 68 ps for a lattice with a reduced trap distance of $v = 0.95$ to 459 ps for a lattice characterized by $v = 1.05$. While this trend is expected with increasing confinement under cw laser excitation, an increase by almost one order of magnitude in coherence time is intriguing. The temporal coherence length for $v = 1.05$ on the order of 0.5 ns was measured at an excitation power of $P = 5\ P_{th}$, well above threshold. Typical coherence times reported in literature vary around 50 ps [30] and 150 ps [46] under non-resonant laser excitation. This underlines that localization due to geometrical frustration is an excellent tool to implement a lattice of highly coherent polariton lasers. For non-resonantly excited, high-quality systems the major source of decoherence is the presence of an excitonic reservoir. By avoiding the excitation of a reservoir, which can be achieved for example by using an Optical Parametric Oscillator configuration, comparable coherence times of 0.5 ns have been achieved [47]. More recently, *A. Askitopoulos et al.* have studied the evolution of temporal coherence in polariton condensates using an annulus pump spot of increasing size, spatially separating reservoir and condensate, and found a "giant increase of temporal coherence" [48]. Interestingly, in our pump geometry we realize the spatial separation of reservoir and condensate in the inverse manner, with highest reservoir density being in the CLS center and the polariton lasing stemming from the six surrounding lobes (c.f. Fig. 5 (b)). By this means, when increasing the lattice distance expressed by $v$ while keeping the pump spot size exactly the same, the spatial reservoir-condensate separation increases as well and the "giant increase of temporal coherence" is reproduced in a flatband laser system. Lastly, it should be mentioned that our platform of Kagome flatband states can be realistically driven by electrical injection [49] representing a vital advance towards novel implementations of topological photonic schemes [50] and topological lasers [51].

**Conclusions**

We have, for the first time, experimentally studied flatbands in polaritonic Kagome lattices and have demonstrated selective condensation into the latter. Furthermore, we have studied the spatial and temporal coherence properties as well as the second-order coherence properties of polariton flatband lasers in great detail. in particular, we find a significant increase in coherence time due to the strong localization of flatband states as a result of the geometrical frustration combined with an efficient way to spatially separate reservoir and flatband condensate. These results lead the way towards in depth studies of interacting quantum fluids of light in a flatband system, linked to topological properties, strong non-linearities, open driven-dissipative systems and flatband lasers.

**Acknowledgements**

T.H.H., C.K., J.B., P.G., J.M., M.E., C.S., S.H., and S.K. acknowledge financial support by the German Research Foundation (DFG) through the Würzburg-Dresden Cluster of Excellence on Complexity and Topology in Quantum Matter "*ct.qmat*" (EXC 2147, project-id 39085490).  S.K., J.B., U.P., and O.A.E. acknowledge support by the German Research Foundation (DFG) within project KL2431/2-1 and  PE523/18-1. S.H. is furthermore grateful for support within the EPSRC Hybrid Polaritonics Grant (Grant No. EP/M025330/1).  T.H.H. and S.H. acknowledge funding by the doctoral training program "Elitenetzwerk Bayern". T.H.H. acknowledges support by the German Academic Scholarship Foundation.

Supporting Information for

# Kagome Flatbands for Coherent Exciton-Polariton Lasing

*T. H. Harder,[1,*] O. A. Egorov,[2] C. Krause,[1] J. Beierlein,[1] P. Gagel,[1] M. Emmerling,[1] C. Schneider,[1,3] U. Peschel,[2] S. Höfling[1,4], and S. Klembt[1,#]*

[1]Technische Physik, Wilhelm-Conrad-Röntgen-Research Center for Complex Material Systems, and Würzburg-Dresden Cluster of Excellence ct.qmat, Universität Würzburg, Am Hubland, D-97074 Würzburg, Germany

[2]Institute of Condensed Matter Theory and Solid State Optics, Abbe Center of Photonics, Friedrich-Schiller-Universität Jena, Max-Wien-Platz 1, D-07743 Jena, Germany

[3]Institute of Physics, University of Oldenburg, D-26129 Oldenburg, Germany

[4]SUPA, School of Physics and Astronomy, University of St. Andrews, KY 16 9SS, United Kingdom

[*]tristan.harder@uni-wuerzburg.de

[#]sebastian.klembt@uni-wuerzburg.de

**Experimental methods**

The polaritonic Kagome lattices were studied using a Fourier spectroscopy setup where the sample is mounted in a liquid Helium flow cryostat operating at 4K. The sample is excited using a continuous wave laser tuned to the first high energy Bragg minimum of the stopband at 1.691eV that is focused onto the sample using a microscope objective with a numerical aperture of 0.4. For the second order correlation measurements, a pulsed laser with a pulse length of 10 ps and a repetition rate of 82MHz was used. The resulting photoluminescence (PL) emission is collected through the same objective and imaged onto the entrance slit of a Cherny-Turner spectrometer equipped with a CCD camera. Depending on the lens configuration used to detect the PL signal, either a real space image of the sample or an image of the backfocal plane of the objective, i.e. a Fourier space image, can be obtained. By scanning the image across the entrance slit of the spectrometer, the full information of $E(x,y)$ or $E(k_x,k_y)$ can be acquired. For first order coherence measurements, a Michelson as well as a Mach-Zehnder interferometer were used. Both interferometers consist of a beam splitter that divides the image into two branches. In the Michelson interferometer, the two branches are equipped with a mirror and a retroreflector, respectively, to superimpose the two images onto a CCD camera. Moving the mirror gives control over the delay between the two branches. In the Mach-Zehnder interferometer, the two images are recombined in a second beam splitter, thus allowing to place a telescope in one of the two arms to magnify a part of the image that can be used as a phase reference. For the second order coherence measurements, a fiber-based Hanbury Brown-Twiss interferometer equipped with avalanche photodiodes for single photon counting was used. Furthermore, the position of the cryostat was actively stabilized to ensure consistent excitation conditions throughout the long integration times.

**Theoretical methods**

For the calculation of exciton-polariton condensation dynamics in Kagome lattices, we use a generalized approach based on the Gross-Pitaevskii (GP) model [1]

$$i\hbar\frac{\partial \Psi(r,t)}{\partial t} = \left[-\frac{\hbar^2}{2m_{\text{eff}}}\nabla^2 - \frac{i\hbar\gamma_C}{2} + V_{\text{ext}}(r) + g_C(r)|\Psi(r,t)|^2 \right.$$

$$\left. + \left(g_R(r) + \frac{i\hbar R(r)}{2}\right)n_R(r,t)\right]\Psi(r,t) + i\hbar\frac{d\Psi_{\text{st}}(r,t)}{dt}$$

$$\frac{\partial n_R(r,t)}{\partial t} = -(\gamma_R + R(r)|\Psi(r,t)|^2)n_R(r,t) + P(r)$$

where $\Psi(r,t)$ is a collective wave function of lower-branch polaritons in the plane of the microcavity with $r \equiv \{x,y\}$, and $n_R(r,t)$ is a high-energy exciton reservoir density. We use realistic experimental parameters: Rabi splitting $\hbar\Omega = 9.91 meV$, photon-exciton detuning $\hbar(\omega_{\text{cavity}} - \omega_{\text{exciton}}) \equiv \hbar\Delta_0 = -17.0 meV$, and the effective mass of intracavity photons $m_C = 34.3 \cdot 10^{-6} m_e$ with the free electron mass $m_e$. Then, the effective mass of lower-branch polaritons can be estimated by $m_{\text{eff}} = 2m_C\sqrt{\Delta_0^2 + \Omega^2}/(\sqrt{\Delta_0^2 + \Omega^2} - \Delta_0)$. The effective trapping potential for polaritons is given by the expression $V_{\text{ext}}(r) = \frac{1}{2}(V(r) + \sqrt{(\Delta_0 - V(r))^2 + \Omega^2} - \sqrt{\Delta_0^2 + \Omega^2}$, where the external potential for the intracavity photons is determined by the Kagome lattice consisting of the mesas in the form of super Gauss $V(r) = \sum_i V_0 \exp\left(-\frac{|r-r_i|^{10}}{d^{10}}\right)$ with the potential depth $V_0 = 8.0 meV$ and mesa diameters $d = 2\mu m$ (centered at $r_i$).

Since the depth of the external potential is comparable with the Rabi splitting, the content of the photonic and excitonic components in polaritons can no longer be considered as spatially homogeneous. To take into account the spatial variation, we scale the key system parameters, such as stimulated scattering $R(\mathbf{r})$ and polariton-polariton interaction $g_C(\mathbf{r})$ and $g_R(\mathbf{r})$, in accordance with the local fraction of the excitonic component by rescaling them with the Hopfield coefficients:

$$|C(\mathbf{r})|^2 = \frac{1}{2}\left(1 - \frac{\Delta_0 - V(\mathbf{r})}{\sqrt{(\Delta_0 - V(\mathbf{r}))^2 + \Omega^2}}\right), \quad |X(\mathbf{r})|^2 = 1 - |C(\mathbf{r})|^2.$$

Then, the stimulated scattering from the reservoir to (coherent) polaritons becomes $R(\mathbf{r}) = R_0|X(\mathbf{r})|^2$ with the fitting parameter $\hbar R_0 = 0.01 meV\, \mu m^2$. The strengths of polariton-polariton and polariton-reservoir interactions are characterized by the expressions $g_C(\mathbf{r}) = g|X(\mathbf{r})|^4$ and $g_R(\mathbf{r}) = g|X(\mathbf{r})|^2$, respectively, where $\hbar g = 0.05 meV\, \mu m^2$ describes the effective interaction between bare excitons. The constants $\hbar\gamma_C = 0.1 meV$ and $\hbar\gamma_R = 0.2 meV$ denote the decay rates of condensed polaritons and reservoir, respectively. These losses are compensated by an external off-resonant time-independent optical pump with the injection rate $P(\mathbf{r})$.

The model also accounts for fluctuations of the condensate. The corresponding terms are derived within the truncated Wigner approximation [2,3], giving $d\Psi_{st}(\mathbf{r}_l) = \sqrt{(\gamma_C + R(\mathbf{r}_l)n_R(\mathbf{r}_l))/(4\delta x \delta y)}\, dW_l$. Here, $dW_l$ is a Gaussian random variable characterized by the correlation functions $\langle W_l^* dW_j \rangle = 2\delta_{l,j} dt$ and $\langle dW_l dW_j \rangle = 0$ where $l, j$ are discretization indices for the spatial coordinate $\mathbf{r} \equiv \{x, y\}$.